\documentclass[twocolumn,10pt,showpacs,showkeys,nofootinbib]{revtex4-1}

\usepackage{graphicx}

\usepackage{amssymb}

\RequirePackage{amssymb}
\RequirePackage{amsmath}
\RequirePackage{graphicx}
\RequirePackage{float}
\RequirePackage{helvet}
\RequirePackage{upgreek}
\RequirePackage{ae}
\RequirePackage{pstricks}
\RequirePackage{revsymb}

\begin{document}

\title{Pion-Field Theoretical Description of the $\Updelta^{++}(1236)$-Resonance without QCD}
\author{Helmut Jahn}\email{h.o.jahn@t-online.de}
\affiliation{Faculty of Physics, University of Karlsruhe, 76131 Karlsruhe, Germany}

\begin{abstract}
\noindent Excited states of the nucleon were predicted by Wentzel \citep{Wentzel} already in 1940. Wentzel
showed that such kind of states can be derived from models of meson field theory if they are treated above a
certain coupling strength. In particular for a pseudo-scalar meson field theory such kind of states were
derived by Pauli and Dancoff \citep{Pauli&Dancoff} in 1942 with equal amount of spin and isospin. They can be understood
as arising from excitation of the pionic cloud of the nucleon by which the main part of nuclear forces also is coming
out. The excitation of these states also is the mechanism by which the pion-nucleon resonance scattering is originated
which firstly was experimentally discovered by Fermi \citep{Fermi} in 1955. This connection was confirmed by Wentzel
in 1975 on occasion of his decoration with the Max-Planck-Medal of the German Physical Society.

\noindent On the other hand there has been the development of the particle physics leading to the concept of the quark
model by Gell-Mann and Zweig. Attempts to incorporate the excited nucleon states of
Wentzel, Pauli \& Dancoff and Fermi into the quark model show that this would require
to introduce a new degree of freedom to which the name ``colour'' has been given and which is not needed in the
case of the excited nucleon states of Wentzel and Pauli \& Dancoff. Moreover it has been stated
by Thomas Walcher \citep{Walcher} that simple observables of the hadrons as mass, spin and magnetic moment cannot be derived
from the quantum colour dynamics (QCD) and the same is true for the nuclear forces. Thus genuine field theoretical
methods of strong coupling had to be developed in order to be able to derive the pion-nucleon resonance scattering from a
model of pion field.

\noindent In the following contribution a completion of the method of Wentzel and Pauli \& Dancoff
is sketched following a work of the author \citep{Jahn} and treating the pion-nucleon resonance scattering. By means of a canonical and
nonlinear transformation of variables the original meson field is split into a free meson part and a self field part of
the nucleon. The transformation is defined such that the interaction between the compound nucleon system thus arising and
the free mesons vanish in the case of infinitely strong coupling. The solution of this field splitting problem is
presented and the problem of calculating the pion-nucleon resonance scattering from this strong-coupling theory is treated
on this basis. Numerical results obtained from such a calculation are discussed. For the sake of simplicity the symmetric
pseudoscalar fixed-and-extended-source meson field theory is used and found that the lowest pion-nucleon resonance
scattering is quite well reproduced with essentially the same parameters necessary to reproduce the nucleon-nucleon-potential.
The present investigation will serve to guide more extended considerations in this direction.
\end{abstract}


\maketitle

\section{The model}

In present days it uses to be forgotten that the first theoretical prediction of excited states of the nucleon was
presented by Wentzel \citep{Wentzel} in 1940 and later on in 1942 by Pauli and Dancoff \citep{Pauli&Dancoff}, Houriet
\citep{Houriet} and by Maki, Sato and Tomonaga \citep{Maki&Sato&Tomonaga} and furthermore extended by Tolar \citep{Tolar},
Doebner \citep{Doebner} and the author \citep{Jahn}.
The pion field producing nuclear forces is described according to Pauli and Dancoff by the Hamiltonian
\begin{align}
H&=\frac{1}{2} \sum_{\rho} \int \left(\pi_{\rho}^2(r)+\varphi_{\rho}(r)(-\triangle+\kappa^2)\varphi_{\rho}(r)\right)\,\mathrm{d}r \notag \\
&\quad\,+\frac{g(2\pi)^{\frac{1}{2}}}{\kappa} \sum_{\rho} \tau_{\rho} \int K(r)(\sigma \cdot \nabla)\varphi_{\rho}\,\mathrm{d}V\,,
\end{align}
with the canonical conjugate hermitian field operators $\varphi_{\rho}(r)$ and $\pi_{\rho}(r)$ of the
pion field of mass $g$, of which the commutation relations are
\begin{align}
&\mathrm{i}[\pi_{\rho'}(r'),\varphi_{\rho}(r)]=\delta_{\rho'\rho}\delta(r'-r) \text{ and } \notag \\
&[\pi_{\rho'}(r'),\pi_{\rho}(r)]=[\varphi_{\rho'}(r'),\varphi_{\rho}(r)]=0\,.
\end{align}
In momentum space may this be written according to
\begin{align}
H&=\frac{1}{2} \sum_{\rho} \int \left(p^{*}_{\rho k}p_{\rho k}+\epsilon^2 q^{*}_{\rho k}q_{\rho k}\right)\,\mathrm{d}k \notag \\
&\quad\,+\frac{\mathrm{i}g}{2\pi\kappa} \sum_{i,\rho} \sigma_i \tau_{\rho} \int k_i v(k)q_{\rho k}\,\mathrm{d}k\,,
\end{align}
with $i$, $\rho=1$, 2, 3, where
\begin{align}
&\mathrm{i}[p_{\rho'k'},q_{\rho k}]=\delta_{\rho'\rho}\delta(k'-k), \notag \\
&[p_{\rho'k'},p_{\rho k}]=[q_{\rho'k'},q_{\rho k}]=0\,,
\end{align}
and
\begin{equation}
\epsilon^2=k^2+\kappa^2\,,
\end{equation}
\begin{equation}
q^{*}_{sk}=q_{s-k},\, p^{*}_{sk}=p_{s-k}\,,
\end{equation}
$\sigma_i$ and $\tau_{\rho}$ are the spin and isospin operators of the source. Further
\begin{equation}
v(k)=\int K(\kappa)\exp(\mathrm{i}k \cdot \kappa)\,\mathrm{d}\kappa \text{ and } v(0)=1\,.
\end{equation}
Also
\begin{equation}
K(\kappa) \mapsto \delta(\kappa),\,v(k)\mapsto 1\,.
\end{equation}
The total spin operator is
\begin{equation}
\mathcal{J}_{\rho\sigma}=\int (q_{\rho k}p_{\sigma k}-q_{\sigma k}p_{\rho k})\,\mathrm{d}k+\frac{\tau_{\rho\sigma}}{2}\,,
\end{equation}
and the total angular momentum
\begin{equation}
\mathcal{G}_{ij}=-\sum_{\rho} \int p_{\rho k}\left(\frac{\partial}{\partial k_i}-\frac{\partial}{\partial k_j}\right)q_{\rho k}\,\mathrm{d}k+\frac{\sigma_{ij}}{2}\,.
\end{equation}
The creation operators of total plane wave pions are
\begin{equation}
\hat{a}^{*}_{\rho k}=\left(\frac{\epsilon}{2}\right)^{\frac{1}{2}}q_{\rho k}-(2\epsilon)^{-\frac{1}{2}}\mathrm{i}p_{\rho-k}\,.
\end{equation}

\section{Definition of the method}

A field theoretical model which consists of a meson field with a given fixed or moving source is usually given by a
Hamiltonian in the form
\begin{equation}
H=H_0+H^c\,,
\end{equation}
where $H_0$ is the uncoupled part and $H^c$ is the coupling term. The eigenvalues problem is
\begin{equation}
H\chi_N=E_N \chi_N\,.
\end{equation}
A non-weak coupling method can be defined as follows. Find a new approximate Hamiltonian $H^0$ in such a way that the
remaining part
\begin{equation}
H^w=H-H^0\,,
\end{equation}
of the total Hamiltonian $H$ can be treated as a perturbation for non-weak coupling. Consequently we have three kinds
of states: the eigenstates $\chi_N$ of the total Hamiltonian $H$, the eigenstates $\hat{\theta}_N$ of the uncoupled
Hamiltonian $H_0$ with
\begin{equation}
H_0\hat{\theta}_N=\hat{E}_N\hat{\theta}_N\,,
\end{equation}
and the eigenstates $\theta_N$ of the new approximate Hamiltonian $H^0$ with
\begin{equation}
H^0\theta_N=E_N^0\theta_N\,.
\end{equation}
All three systems of states of course contain no-meson states. The no-meson state $\chi_0$ of the first system given by
\begin{equation}
H\chi_0=E\chi_0\,,
\end{equation}
is the exact physical nucleon state with its meson cloud around. The no-meson state $\hat{\theta}_0$ of the second system
simply is the bare nucleon state characterized by
\begin{equation}
H_0\hat{\theta}_0=\hat{E}\hat{\theta}_0\,.
\end{equation}
The no-meson states $\theta_{0,T}$ of the new approximate system given by
\begin{equation}
H^0\theta_{0,T}=E^0_T\theta_{0,T}\,,
\end{equation}
include besides the approximate physical nucleon state also excited nucleon states with higher spin and isospin. These
are the so-called isobars, which first where discovered by Wentzel \citep{Wentzel}. By means of the new interaction therm $H^w$ the
isobars can decay into mesons and the physical nucleon ground state such giving the well known resonance behaviour of the
pion-nucleon-interaction. The emission and absorption of mesons with momentum $k$ and isospin vector component $\rho$ can
be described by meson-creation and destruction operators $a^{*}_{\rho k}$, $a_{\rho k}$ acting in the system of the new
approximate states $\theta_N$ according to
\begin{equation}
\theta_N^{\pm}=\frac{1}{\sqrt{N!}}a^{\pm *}_{\rho_1 k_1} \hdots a^{\pm *}_{\rho_N k_N}\theta_{0,T}\,.
\end{equation}
The excitation and decay of the isobars can be described by operators $\mathcal{O}^{*}$, $\mathcal{O}$ acting only on the
isobaric quantum number $T$ of the isobaric operator $\mathcal{T}$. Correspondingly there are of course meson creation and
destruction operators acting in the system of the uncoupled states according to
\begin{equation}
\hat{\theta}^{\pm}_N=\frac{1}{\sqrt{N!}}\hat{a}^{\pm *}_{\rho_1 k_1} \hdots \hat{a}^{\pm *}_{\rho_N k_N}\hat{\theta}_0\,.
\end{equation}
If we then are able to find an operator relationship
\begin{equation}
\hat{a}=\hat{a}(a^{*},a,\mathcal{O}^{*},\mathcal{O},\mathcal{T})\,,
\end{equation}
we then can construct the new approximate Hamiltonian $H^0$ and the new interaction term $H^w$ by inserting this operator
relationship into the new form of the total Hamiltonian $H$. According to Coester \citep{Coester} the $S$-matrix-element
$S_{N'N}=(\chi^-_{N'},\chi^+_N)$ for the meson-nucleon interaction in the representation of the new approximate states
$\theta^{\pm}_N$ is then given by
\begin{align}
S_{N'N}&=(\theta^-_{N'},\theta^+_{N})-\frac{2\pi\mathrm{i}}{Z}\delta(E_{N'}-E_N)[(\theta^-_{N'},H^w\theta^+_N) \notag \\
&\quad\,+(\theta^-_{N'},H^w(E_N-H+\mathrm{i}\varepsilon)^{-1}H^w\theta^+_N)]\,,
\end{align}
where
\begin{equation}
Z=(\chi_0,\theta_{0,1/2})^2\,,
\end{equation}
and can be calculated according to
\begin{align}
Z^{-1}-1&=(\theta_{0,1/2},H^w(1-\Lambda_{1/2}) \notag \\
&\quad\,\times [H^0+(1-\Lambda_{1/2})H^w-E]^{-2} \notag \\
&\quad\,\times(1-\Lambda_{1/2})H^w\theta_{0,1/2})\,.
\end{align}
$\Lambda_{1/2}$ is the projection operator into the new approximate ground state $\theta_{0,1/2}$ defined by
\begin{equation}
\Lambda_{1/2}\theta_{0,1/2}=\theta_{0,1/2}\,.
\end{equation}
The above equations show that a proper approximation is only obtained if $Z^{-1}$ can be calculated by treating $H^w$ as
a perturbation. An indication for this being the case would be the smallness of the lowest order perturbation theoretical
expression
\begin{align}
{Z^{(1)}}^{-1}-1&=(\theta_{0,1/2},H^w(1-\Lambda_{1/2})[H^0-E]^{-2} \notag \\
&\quad\,\times(1-\Lambda_{1/2})H^w\theta_{0,1/2}) \ll 1\,.
\end{align}
This relation would provide a criterium for finding a proper operator relationship for $\hat{a}$. The usual weak-coupling
renormalization constant
\begin{equation}
\overline{Z}^{\frac{1}{2}}=(\chi_0,\hat{\theta}_{0,1/2})\,,
\end{equation}
does not occur in Eq.~(23). It is not at all needed anymore and therefore can have any value smaller than 1 including
zero. In order to exhibit the pion-nucleon resonance scattering caused by the 3/2-isobar $\theta_{0, 3/2}$ we used
according to Coester a projection operator $\Lambda_{3/2}$ into $\theta_{0, 3/2}$ defined by
\begin{equation}
\Lambda_{3/2}\theta_{0, 3/2}=\theta_{0, 3/2}\,.
\end{equation}
We define
\begin{equation}
v'=(1-\Lambda_{3/2})H^w(1-\Lambda_{3/2})\,,
\end{equation}
and
\begin{equation}
v=(1-\Lambda_{3/2})H^w\Lambda_{3/2}+\Lambda_{3/2}H^w(1-\Lambda_{3/2})\,,
\end{equation}
with
\begin{equation}
\|v'(E_N-H^0+\mathrm{i}\varepsilon)^{-1}v'\| \ll 1\,,
\end{equation}
the last term in Eq.~(23) can be rewritten
\begin{align}
(\theta^-_{N'},H^w&(E_N-H+\mathrm{i}\varepsilon)^{-1}H^w \theta^+_N) \notag \\
&=(\theta^-_N,v(E_N-H^0-\nabla)^{-1}v\theta^+_N) \notag \\
&=\frac{(\theta^-_{N'},v\theta_{0, 3/2})(\theta_{0, 3/2}, v\theta^+_N)}{E_N-(\theta_{0, 3/2}, H^0\theta_{0, 3/2})-\nabla}\,,
\end{align}
where
\begin{align}
\nabla&=\Lambda_{3/2}v(E_N-H^0+\mathrm{i}\varepsilon)^{-1} v\Lambda_{3/2} \notag \\
&=4\pi \int\limits_{0}^{\infty} \frac{|(\theta_{0, 3/2},v\theta_{N'})|^2}{\epsilon-\epsilon'+\mathrm{i}\varepsilon}k'\epsilon'\,\mathrm{d}\epsilon'\,.
\end{align}
With Eq.~(33) the resonance behaviour is exhibited. Thus $\| v'(E_N-H^0+\mathrm{i}\varepsilon)^{-1}v'\|\ll 1$ provides
another criterium for finding a proper relationship for $\hat{a}$ to describe the resonance behaviour of the scattering
process. The operator substitution for $\hat{a}$ is
\begin{equation}
q_{\rho k}=q'_{\rho k}+\sum_i u_{k_i}s_{i\rho}\,;
\end{equation}
\begin{equation}
p_{\rho k}=p'_{\rho k}+\sum_i \frac{u_{k_i}p_{i\rho}}{\int |u_{k}|^2\,\mathrm{d}k}\,;
\end{equation}
with
\begin{equation}
q'_{\rho \mathbf{k}}=\sum_{\rho_0} \int (2\epsilon)^{-\frac{1}{2}}({v^{\mathbf{kk_0}}_{\rho\rho_0}}^{*}a_{\rho_0 \mathbf{k_0}}+v^{\mathbf{kk_0}}_{\rho\rho_0}a^{*}_{\rho_0\mathbf{k_0}})\,\mathrm{d}\mathbf{k_0}\,,
\label{eq:definition-qrhokprime}
\end{equation}
\begin{equation}
p'_{\rho \mathbf{k}}=\mathrm{i}\sum_{\rho_0} \int \left(\frac{\epsilon}{2}\right)^{\frac{1}{2}}(v^{\mathbf{kk_0}}_{\rho\rho_0} a^{*}_{\rho_0\mathbf{k_0}}-{v_{\rho\rho_0}^{\mathbf{kk_0}}}^{*}a_{\rho_0\mathbf{k_0}})\,\mathrm{d}\mathbf{k_0}\,,
\end{equation}
and
\begin{equation}
A_{i\rho}=\sum_{\sigma} s_{i\sigma}(J'_{\rho\sigma}-J_{\rho\sigma})\,,
\end{equation}
with the total isospin components
\begin{align}
&J_{\rho\sigma}=\int (q_{\rho k}p_{\sigma k}-q_{\sigma k}p_{\rho k})\,\mathrm{d}k+\frac{\tau_{\rho\sigma}}{2}, \notag \\
&J'_{\rho\sigma}=\int (q'_{\rho k}p'_{\sigma k}-q'_{\sigma k}p'_{\rho k})\,\mathrm{d}k\,,
\end{align}
and $\mathcal{J}_{\rho\sigma}$ and $\mathcal{G}_{\rho\sigma}$ as before. Further
\begin{equation}
q=\sum_{i,\sigma} s_{i\sigma}\frac{\int u_{k_j}q'_{\sigma k}\,\mathrm{d}k}{\int |u_{k_i}|^2\,\mathrm{d}k}\,.
\end{equation}
Following Eq.~(16) and (35) we obtain
\begin{equation}
H^0=H^{0,f}+H^{0,c}\,,
\end{equation}
with
\begin{equation}
H^{0,f}=\int \epsilon_0 \langle a^{*}_{\rho_0k_0}a_{\rho_0 k_0}\rangle \,\mathrm{d}k_0\,,
\end{equation}
and
\begin{align}
H^{0,c}&=\frac{1}{2} \sum_{\rho} \int \left({p^{e}_{\rho k}}^{*}p^e_{\rho k}+\epsilon^2 {q^{e}_{\rho k}}^{*}q^{e}_{\rho k}\right)\,\mathrm{d}k \notag \\
&\quad\,+\frac{\mathrm{i}g}{2\pi\kappa} \sum_{i,\rho} \sigma_i\tau_{\rho} \int k_i v(k)q^e_{\rho k}\,\mathrm{d}k\,,
\end{align}
where
\begin{equation}
q^e_{\rho k}=\sum_i u_{k_i}s_{i\rho}, \qquad p^e_{\rho k}=\sum_i\frac{u_{k_i}p_{i\rho}}{\int |u_{k_i}|^2\,\mathrm{d}k}\,,
\end{equation}
with the compound nucleon variables $s_{i\rho}$
\begin{widetext}
\begin{equation}
\begin{pmatrix}
\cos\theta\cos\phi\cos\psi-\sin\phi\sin\psi & \cos\theta\cos\phi\sin\psi+\sin\phi\cos\psi & -\sin\theta\cos\phi \\
-\cos\theta\sin\phi\cos\psi-\cos\phi\sin\psi & -\cos\theta\sin\phi\sin\psi+\cos\phi\cos\psi & \sin\theta\sin\phi \\
\sin\theta\cos\psi & \sin\theta\sin\psi & \cos\theta \\
\end{pmatrix}\,,
\end{equation}
\end{widetext}
and
\begin{equation}
p_{i\rho}=-\sum_{\sigma} \langle s_{i\sigma}\mathcal{T}^{e}_{\rho\sigma}\rangle=-\sum_j \langle s_{j\rho}\mathcal{L}^e_{ij}\rangle\,,
\end{equation}
where $\langle\bullet\rangle$ means symmetrized product, and
\begin{align}
&\mathcal{L}^e_{12}=\frac{1}{\mathrm{i}}\frac{\partial}{\partial\phi}, \notag \\
&\mathcal{L}^e_{23}=-\sin\phi \frac{1}{\mathrm{i}}\frac{\partial}{\partial\theta}+\frac{\cos\phi}{\sin\theta}\left(\frac{1}{\mathrm{i}}\frac{\partial}{\partial\psi}-\cos\theta \frac{1}{\mathrm{i}}\frac{\partial}{\partial\phi}\right), \notag \\
&\hdots\,,
\end{align}
are the angular momentum operators and in addition
\begin{align}
&\mathcal{T}_{12}^e=\frac{1}{\mathrm{i}}\frac{\partial}{\partial\psi}, \notag \\
&\mathcal{T}^{e}_{23}=-\sin\psi \frac{1}{\mathrm{i}}\frac{\partial}{\partial\theta}+\frac{\cos\psi}{\sin\theta}\left(\frac{1}{\mathrm{i}}\frac{\partial}{\partial\phi}-\cos\theta\frac{1}{\mathrm{i}}\frac{\partial}{\partial\psi}\right), \notag \\
&\hdots\,,
\end{align}
are the isospin operators of the composed nucleon. These variables can be interpreted as describing a symmetric top of which
the angular momentum is described in the space-fixed system whilst the isospin is described in the body-fixed system with
equal amount of quantum numbers according to
\begin{equation}
{\mathcal{L}^{e}}^2={\mathcal{T}^{e}}^{2} \text{ such } L(L+1)=T(T+1)\,.
\end{equation}
where $L$ and $T$ are the eigenvalues of the compound nucleon spin and isospin and where $s_{i\rho}$ transforms from the
space-fixed system to the body-fixed system. This model is the result of the strong-coupling method of Wentzel \citep{Wentzel}, Pauli \&
Dancoff \citep{Pauli&Dancoff} and the author \citep{Jahn}, according to which the complete operator relationship will be met with their coefficients
$u_{k_i}$ and $v_{\rho\rho_0}^{kk_0}$. With Eq.~(3), $H^{0,f}$ and Eq.~(43) the authors' formulation of the strong
coupling method has been stated as follows. With
\begin{equation}
H^w=H-(H^{0,f}+H^{0,c})\,,
\end{equation}
the perturbation series
\begin{equation}
f_H(n)=\delta_{Nn}+(n|H^w|N)/(E_N-\epsilon_n)+\hdots\,,
\end{equation}
based on the expansion
\begin{equation}
\chi_N=\sum_n f_N(n)\theta_n\,,
\end{equation}
of the eigenstates of $H$ after the eigenstates of $H^{0,f}+H^{0,c}$ according to $H\chi_N=E_N\chi_N$ and $H^0\theta_N=E_N^0\theta_N$
and with fixed distance between the energies $E_N$ and the ground state energy $E_0$ it then gets
\begin{equation}
f_N(n) \mapsto \delta_{Nn} \text{ for } g\mapsto \infty\,.
\end{equation}
But inserting the operators substitution Eq.~(35) and taking into account $H^0=H^{0,f}+H^{0,c}$ and Eq.~(43) we
obtain
\begin{align}
H^w&=\frac{1}{2}\sum_{\rho}\int \left({p'_{\rho k}}^{*}p'_{\rho k}+\epsilon^2 {q'_{\rho k}}^{*}q'_{\rho k}\right)\,\mathrm{d}k-H^{0,f} \notag \\
&\quad\,+\sum_{i,\rho} \int \left(2s_{i\rho}u_{k_i}\epsilon^2+\frac{\mathrm{i}g}{2\pi\kappa} \sigma_i \tau_{\rho} k_i v(k)\right)q'_{\rho k}\,\mathrm{d}k\,.
\end{align}

\section{Dynamics of the composed nucleon}

The eigenvalue of the no-meson states of the new approximate system will be obtained according to $H^{0,c}\theta_{0,T}=E^0_T\theta_{0,T}$. We
are getting
\begin{align}
H^{0,c}&=\frac{3}{2 \sum_i \int |u_{k_i}|^2\,\mathrm{d}k}{\mathcal{L}^e}^2+\frac{1}{2}\sum_i \int \epsilon^2|u_{k_i}|^2\,\mathrm{d}k \notag \\
&\quad\,+\frac{\mathrm{i}g}{2\pi\kappa} \sum_i \sigma_i\tau_i \int k_i v(k)u_{k_i}\,\mathrm{d}k\,,
\end{align}
where
\begin{equation}
\tau_i=\sum_{\rho} \tau_{\rho}s_{i\rho}\,,
\end{equation}
is the isospin-component relative to the spin-motion. Thus the coupling term in $H^{0,c}$ formally describes the coupling
of two spins with singlet-triplet-states\footnote{This corresponds to Hund's coupling case of molecular physics, where the
spin of lightning electrons is directed along the rotating molecular axis \citep{Hund}.}, where the singlet-states are the
ground states of the composed nucleon. In the case of strong coupling only the coupling term and the term before it
survive in zero-approximation and the $u_k$-function which minimizes the term becomes
\begin{equation}
u_{k_i}=\frac{\mathrm{i}gk_i v(k)}{2\pi \kappa \epsilon^2}\,.
\end{equation}

\section{Resonance scattering}

We obtain the resonance scattering if we insert the corresponding terms of the precedingly developed formalism into the
expressions Eq.~(33). With the strong coupling solution $u_{k_i}$ we obtain the lowest excitation step from the first
term of Eq.~(56) yielding for the resonance denominator in Eq.~(33) with $H^0\theta_{0,T}=E^0_T\theta_{0,T}$ as
\begin{align}
E_N-(\theta_{0, 3/2}&,H^0\theta_{0, 3/2})-\nabla \notag \\
&=\epsilon-(E^0_{3/2}-E^0_{1/2})-\mathrm{Re}(\nabla)+\frac{\mathrm{i}}{2}\Gamma\,,
\end{align}
where we have put the incident energy $E_N$ into a pion part $\epsilon$ and a part $E^0_{1/2}$ of the nucleon ground state
\begin{equation}
E_N=\epsilon+E_{1/2}^0\,,
\end{equation}
and from the first term of Eq.~(56) we obtain
\begin{equation}
E_{3/2}^0-E_{1/2}^0=\frac{9}{2\sum_i \int |u_{k_i}|^2\,\mathrm{d}k}\,.
\end{equation}
The interaction terms causing this excitation and deexcitation is given by the second term of Eq.~ (55)
to first order in $q$ as considering as the only particle producing and absorbing operators and matrix
elements contributing to Eq.~(33) according to
\begin{equation}
\left\langle 0,\frac{3}{2},\frac{3}{2}\right|v\left|k,\frac{1}{2},\frac{1}{2}\right\rangle
=\left\langle 0,\frac{3}{2},\frac{3}{2}\right|Q\left|k,\frac{1}{2},\frac{1}{2}\right\rangle\,,
\end{equation}
with
\begin{equation}
Q=\sum_{i,\rho} \int \left(2s_{i\rho}u_{k_i}\varepsilon^2+\frac{\mathrm{i}g}{2\pi\kappa}\sigma_i\tau_{\rho}k_i v(k)\right)q'_{\rho k}\,\mathrm{d}k\,,
\end{equation}
hence a Breit-Wigner resonance according to Feshbach \cite{Coester}.
Thus the resonance energy can be calculated from the difference $E^0_{3/2}-E^0_{1/2}$ and the level shift with resonance
width can be calculated from Eq.~(33) with $\nabla$ and Eq.~(61). With Eq.~(57) and Eq.~(60) we obtain
$E^0_{3/2}-E^0_{1/2}=9\pi \kappa^2/2K g^2$ which is exactly in agreement with the result of Pauli and Dancoff \citep{Pauli&Dancoff}
for the resonance energy of the lowest pion-nucleon-resonance $=296\,\mathrm{MeV}$ with pion restmass
$\kappa c=140\,\mathrm{MeV}$ and cut-off momentum $K$ corresponding to the nucleon restmass
$Kc=940\,\mathrm{MeV}$ and $g=1$. Experimental data are from \cite{Musiol}.

\begin{figure}[h]
\centering
\includegraphics[scale=0.50]{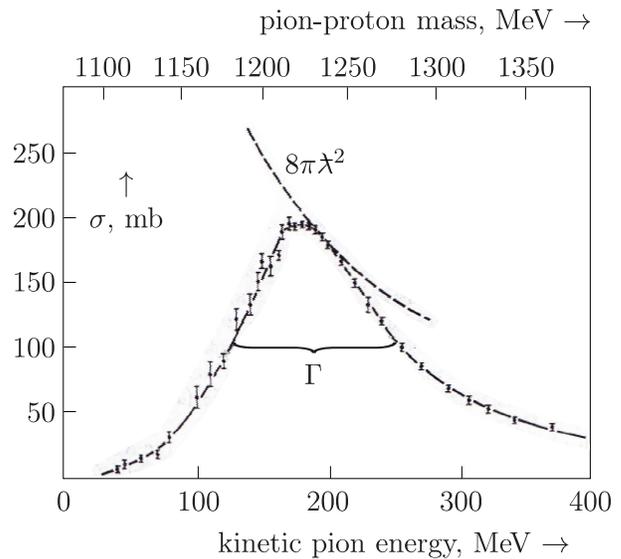}
\caption{The cross section of $\uppi^+\mathrm{p}$ elastic scattering is shown in the vicinity of the $\Updelta^{++}(1232)$ resonance, whose width is $\Gamma=120\,\mathrm{MeV}$. The maximum elastic cross section $8\pi\lambdabar^2$ is reached above the peak of the resonance, where $\lambdabar=\hbar/p$ is the reduced de-Broglie wavelength.}
\label{fig:resonance}
\end{figure}

Evaluation of the width of the resonance according to Eqs.~(33), (34) gives
\begin{equation}
\Gamma=2\pi
\left|\left\langle 0,\frac{3}{2},\frac{3}{2}\right|v\left|k,\frac{1}{2},\frac{1}{2}\right\rangle\right|^2\,,
\end{equation}
in agreement with Feshbach's expression (page 163). The explicit result is given by
\begin{equation}
\Gamma=\frac{g^2}{3}\frac{k^3}{\kappa^2}\sum_{i,\rho} |s_{i\rho}|^2\,.
\end{equation}
Using the Pauli approximation,
hence $\sigma_i\tau_{\rho}=-s_{i\rho}$, we obtain:
\begin{equation}
\Gamma=\frac{g^2}{3}892.22\,\mathrm{MeV} \sum_{i,\rho} |s_{i\rho}|^2\,,\quad \sum_{i,\rho} |s_{i\rho}|^2=\frac{1}{3}\,.
\end{equation}
The final result is $\Gamma=99.14\,\mathrm{MeV}$ (compare to the curve in Fig.~\ref{fig:resonance} below that can be found in \cite{Perkins} and has been slightly adapted).

\section*{Acknowledgments}

Valuable discussions and help for getting the manuscript are gratefully acknowledge to A. Anzaldo, H.~D. Doebner,
M. Schreck, T.~H. Seligman and J. Tolar.

\end{document}